\documentclass[conference]{IEEEtran}
\IEEEoverridecommandlockouts

\usepackage{cite}
\usepackage{amsmath,amssymb,amsfonts}
\usepackage{graphicx}
\usepackage{booktabs}
\usepackage{array}
\usepackage{multirow}
\usepackage{tabularx}
\usepackage{makecell}
\usepackage{xcolor}
\usepackage{colortbl}
\usepackage{stfloats}
\usepackage{textcomp}
\usepackage{url}
\usepackage{listings}
\usepackage{balance}
\usepackage[hidelinks]{hyperref}

\lstdefinestyle{plain}{
    basicstyle=\ttfamily\footnotesize,
    breaklines=true,
    columns=fullflexible,
    frame=single,
    framesep=4pt,
    xleftmargin=4pt,
    aboveskip=6pt,
    belowskip=6pt,
}

\definecolor{beforered}{HTML}{B33A3A}
\definecolor{aftergreen}{HTML}{2F8A4F}

\lstdefinestyle{plainBefore}{
    basicstyle=\ttfamily\footnotesize,
    breaklines=true,
    columns=fullflexible,
    frame=single,
    framesep=5pt,
    xleftmargin=5pt,
    aboveskip=2pt,
    belowskip=6pt,
    backgroundcolor=\color{red!4},
    rulecolor=\color{beforered!45},
    framerule=0.6pt,
    moredelim=[is][\bfseries\color{beforered}]{(*}{*)},
}

\lstdefinestyle{plainAfter}{
    basicstyle=\ttfamily\footnotesize,
    breaklines=true,
    columns=fullflexible,
    frame=single,
    framesep=5pt,
    xleftmargin=5pt,
    aboveskip=2pt,
    belowskip=6pt,
    backgroundcolor=\color{green!5},
    rulecolor=\color{aftergreen!45},
    framerule=0.6pt,
}

\newcommand{\valci}[2]{\makecell{#1\\\ci{#2}}}

\begin{document}

\title{MOLOT System Card: Malicious Operational Logic Observation Transformer%
\thanks{Corresponding author: M.~Mitrofanov, False Positive Community, \href{mailto:falseposi@yandex.com}{\nolinkurl{falseposi@yandex.com}}.}%
\thanks{\IEEEauthorrefmark{1}These authors contributed equally to this work.}}

\author{
\IEEEauthorblockN{Daniil Lopatkin\IEEEauthorrefmark{1}}
\IEEEauthorblockA{Data Science \& ML Dept.\\Positive Technologies\\False Positive Community}
\and
\IEEEauthorblockN{Maksim Mitrofanov\IEEEauthorrefmark{1}}
\IEEEauthorblockA{Data Science \& ML Dept.\\Positive Technologies\\False Positive Community}
\and
\IEEEauthorblockN{Stanislav Rakovsky}
\IEEEauthorblockA{Expert Security Center Dept.\\Positive Technologies\\False Positive Community}
\and
\IEEEauthorblockN{Aleksandr Khalikov}
\IEEEauthorblockA{Application Security Dept.\\Positive Technologies\\False Positive Community}
}

\maketitle

\begin{abstract}
MOLOT (Malicious Operational Logic Observation Transformer) is a static malicious-code detection system designed for SAST setup where package metadata, maintainer history, and dynamic execution traces may be unavailable or unreliable. The system represents source code as behavior sequences derived from static call graphs, includes an explanation stage that ranks suspicious behavior activities and maps them back to source-code locations. The approach is evaluated on Python and JavaScript packages from PyPI and npm, compared with open-source detection tools, and validated under product constraints including runtime, memory use, and false-positive rates observed in a real moderation workflow. We also release Open Malicious-code Bench, a public benchmark for reproducible evaluation of malicious-package detection methods. The results show that static behavior-sequence modeling can provide accurate, explainable, and deployable malicious-code detection for modern DevSecOps workflows.
\end{abstract}


\section{Introduction}
\label{sec:intro}

Hackers increasingly rely on embedded malicious code, classified as CWE-506, as a practical way to harm companies, organizations, and software projects. Recent supply-chain incidents illustrate the problem: campaigns around \href{https://docs.litellm.ai/blog/security-update-march-2026}{LiteLLM}~\cite{dholakia2026litellm} and \href{https://securitylabs.datadoghq.com/articles/shai-hulud-2.0-npm-worm/}{Shai-Hulud~2.0}~\cite{tafani2025shaihulud} abused ordinary project artifacts such as \texttt{litellm\_init.pth}, \texttt{setup\_bun.js}, and \texttt{bun\_environment.js} to hide malicious behavior in project source code or build-time logic. These cases show that malicious-package detection cannot be treated only as a registry moderation problem; secure development workflows also need techniques that can analyze code before publication and before execution.

Existing malicious-package detection systems often combine behavioral features with package metadata, maintainer history, registry popularity, dependency context, or dynamic execution traces. These signals are useful for public PyPI and npm moderation, but they are weak or unavailable in static application security testing scenarios: a package may be private, unpublished, freshly vendored, or part of an internal repository. In addition, metadata-heavy approaches can be brittle when an attacker compromises a legitimate account or injects malicious logic into otherwise trusted projects. For SAST, the core signal must therefore come from the source code itself.

MOLOT addresses this setting by representing source code as behavior sequences derived from a static call graph. Building on prior behavior-sequence approaches~\cite{zhang2024killing}, our system focuses on improving the call-chain representation itself: it removes file and method names that can cause overfitting, while preserving security-relevant information from literals and surrounding calls to make the resulting sequences more diverse and behavior-oriented. Expert labeling combined with LLM-assisted annotation also enables MOLOT to move from package-level classification to file-level classification, which reduces pressure on the model context window and makes it possible to detect malicious code in larger projects.

A second design goal of MOLOT is operational usefulness. Detection results must be explainable enough for a security engineer to inspect the code that triggered a decision. For this reason, MOLOT includes an explanation stage based on SHAP over behavior activities and returns the most influential suspicious calls together with source locations. The system is also evaluated not only on curated validation data, but on product-oriented constraints: runtime, memory consumption, and false-positive rates observed in a real moderation workflow.

This paper makes the following contributions:
\begin{itemize}
    \item We present MOLOT, a static behavior-sequence approach for malicious-code detection that is suitable for SAST and supply-chain security scenarios where registry metadata may be unavailable.
    \item We describe the construction of expert labeling and LLM-assisted file-level annotation.
    \item We evaluate model and pipeline modifications that improve detection quality and add an explanation module that links suspicious behavior back to code locations.
    \item We validate the system against open-source tools and in a real-world package moderation workflow, showing that the approach is practical under product constraints.
    \item We release Open Malicious-code Bench, a public benchmark for reproducible comparison of malicious-package detection methods.
\end{itemize}

\textbf{Terminology.} We evaluate MOLOT on Python and JavaScript packages; for brevity we abbreviate JavaScript as JS and treat TypeScript as part of JavaScript, so every later reference to JavaScript or JS also covers TypeScript.

\textbf{Paper Organization.} The rest of the paper is organized as follows. Section~\ref{sec:related} reviews prior work on malicious-package detection, behavior-sequence models, graph-based analysis, dynamic analysis, LLM-assisted labeling, and public benchmarks. Section~\ref{sec:methodology} describes the MOLOT methodology, including call-graph extraction, behavior-sequence generation, file-level labeling, model training, explainability, and performance evaluation. Section~\ref{sec:experiments} reports experimental results, covering the impact of key pipeline modifications, language-specific performance, comparisons with open-source tools, the explainability evaluation, runtime and memory measurements, and observations from a real-world moderation deployment. Section~\ref{sec:limitations} discusses limitations and future work, Section~\ref{sec:data} describes the released benchmark, and Section~\ref{sec:conclusion} concludes the paper.

\section{Related Work}
\label{sec:related}

The problem of detecting malicious packages in the PyPI and npm ecosystems has attracted significant attention in recent years. Existing research generally falls into five categories: behavioral-sequence modeling with BERT, graph-based detection, dynamic analysis, LLM-assisted methods, and benchmark-driven empirical studies. In this section, we review these directions and explain how MOLOT relates to them.

\subsection{Behavioral-Sequence Models}

The main precursor of our approach is CEREBRO~\cite{zhang2024killing}, which first showed that a single BERT model trained on high-level behavioral descriptions can generalize across both PyPI and npm. We reproduced this approach as our baseline and later extended it with the modifications described in Section~\ref{sec:key_mods}.

Ea4mp~\cite{sun2024integrating} builds on the same idea by combining behavioral sequences with package metadata in an ensemble framework. Besides behavioral information, the model uses registry-level signals such as account age and publication history. MOLOT intentionally avoids such metadata. In SAST scenarios, these signals are often unavailable. In moderation pipelines, they can also be unreliable because malicious campaigns may use compromised accounts belonging to legitimate contributors (see Section~\ref{sec:limitations}).

CLAMPD-Net~\cite{iqbal2026clampd} continues this trend toward richer feature integration. Instead of relying solely on transformers, it introduces a multimodal CNN-BiGRU-Transformer architecture that combines API patterns, dependency graphs, and metadata. Overall, these works illustrate a broader trend toward increasingly complex architectures designed to achieve incremental performance gains.

\subsection{Graph-Based Approaches}

Another line of research models package behavior as graphs. SpiderScan~\cite{huang2024spiderscan} represents npm package behavior as a call graph and detects malicious activity by matching against expert-defined subgraphs. Conceptually, this is the closest existing approach to MOLOT.

Both SpiderScan and MOLOT operate at the call-graph level. The key difference is that SpiderScan relies on manually defined rules, whereas MOLOT replaces them with a trainable BERT classifier. This distinction helps explain the F1 gap observed in comparisons with traditional SAST tools (see Section~\ref{sec:opensource}).

\subsection{Dynamic Analysis}

Dynamic analysis offers an alternative to static detection methods. DySec~\cite{mehedi2025dysec} collects 36 runtime features during package installation, including system calls, network activity, and resource consumption, and trains a classifier using these signals.

Although the approach achieves strong detection performance, it requires executing packages inside a sandbox. As a result, it is unsuitable for SAST pipelines that must operate before code execution.

\subsection{LLM-Based Detection and Labeling}

Another important direction is the direct use of LLMs for malicious-package detection. Ibiyo et al.~\cite{ibiyo2025detecting} demonstrate that few-shot prompting enables LLMs to detect malicious code in PyPI packages at the file level with accuracy reaching 97\%.

These results motivate our use of LLMs as ``teachers'' for file-level labeling (see Section~\ref{sec:filelevel}). However, MOLOT delegates final inference to a computationally cheaper BERT model.

Similar findings were later reported by Zahan et al.~\cite{zahan2025leveraging}. Their SecurityAI pipeline achieved 99\% precision and 97\% F1 on a benchmark of 5{,}115 npm packages, substantially outperforming the rule-based CodeQL~\cite{codeql}. These results further support the use of LLM-generated supervision while also highlighting the limitations of rule-based systems such as semgrep~\cite{semgrep} and bandit4mal~\cite{bandit4mal}, which we use as baselines in Section~\ref{sec:opensource}.

\subsection{Benchmarks and Taxonomies}

The benchmark and taxonomy components of MOLOT are most closely related to three previous studies.

Guo et al.~\cite{guo2026understanding} introduced a benchmark containing 13{,}708 npm packages and evaluated 8 tools across 13 configurations. They also proposed a taxonomy covering 11 malicious-behavior classes and 8 evasion techniques. Their methodology closely aligns with our Section~\ref{sec:publicbench} and motivates the publication of an open subset of the MOLOT validation set.

Ryan et al.~\cite{ryan2025unveiling} proposed a taxonomy of 47 indicators of malicious activity at the statement level. Their taxonomy aligns closely with the activity classes used in our file-level labeling procedure.

Finally, Ladisa et al.~\cite{ladisa2023feasibility} studied cross-language transfer between npm and PyPI and demonstrated that such transfer is feasible using language-independent features. Their findings provide theoretical support for our use of a unified BERT model for both Python and JS (see Section~\ref{sec:language}).

\subsection{Positioning of MOLOT}

MOLOT combines ideas from several research directions. Like CEREBRO~\cite{zhang2024killing} and Ea4mp~\cite{sun2024integrating}, it relies on behavioral sequences. Similar to SpiderScan~\cite{huang2024spiderscan}, it operates at the call-graph level. However, unlike multimodal approaches~\cite{sun2024integrating,iqbal2026clampd}, MOLOT avoids metadata features in order to remain applicable in SAST environments.

In contrast to dynamic systems such as DySec~\cite{mehedi2025dysec}, MOLOT preserves a fully static analysis pipeline. At the same time, unlike LLM-centered approaches~\cite{ibiyo2025detecting,zahan2025leveraging}, it uses LLMs only during dataset preparation to scale supervision while keeping inference efficient and deployable.

Finally, the evaluation methodology presented in Section~\ref{sec:publicbench} follows principles similar to those introduced in prior benchmark and taxonomy studies~\cite{guo2026understanding,ryan2025unveiling,ladisa2023feasibility}.

\section{Methodology}
\label{sec:methodology}

Having positioned MOLOT among existing approaches, we now describe its design in detail: how source code is turned into behavior sequences, how those sequences are labeled and classified, and how predictions are explained and validated under product constraints.

\subsection{Approach}
\label{sec:approach}

We turn each project into a textual sequence of code-execution activities, classify those sequences with a BERT-based transformer, and surface per-activity SHAP attributions for the samples the model flags as malicious.

The textual sequence is derived from the project's call graph. Once the transformer assigns a binary verdict, \href{https://shap.readthedocs.io/en/latest/}{SHAP} is used to identify which activities drove a ``malicious'' prediction.

\begin{figure*}[t]
    \centering
    \includegraphics[width=\textwidth]{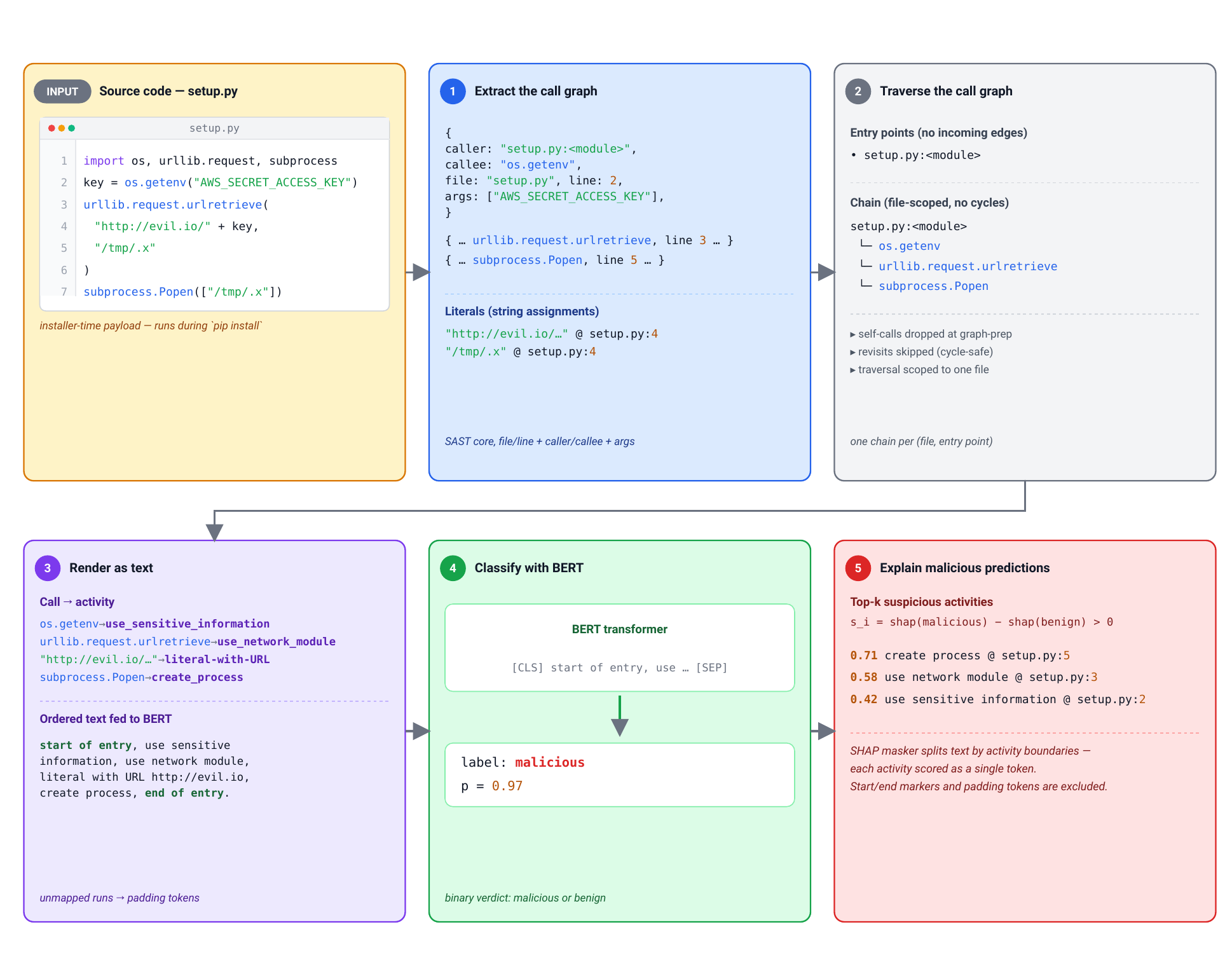}
    \caption{The MOLOT pipeline: call-graph extraction, traversal into activity chains, textual rendering, BERT classification, and SHAP-based explanation.}
    \label{fig:pipeline}
\end{figure*}

The package-processing pipeline (Fig.~\ref{fig:pipeline}) has five stages:

\begin{enumerate}
    \item \textbf{Extract the call graph.}
    The PT Application Inspector SAST engine produces the project's call graph. For every function call we record its location in the code (file and line), its arguments, and the name of the enclosing function whose body contains the call.
    We also capture the literals defined in the code: some call arguments are bound to local variables before they reach the call site, so their concrete values are invisible there. We initially planned to recover such values with data flow analysis, but it inflated project-processing time prohibitively. Instead we record every point where a literal is assigned to a variable, as a separate event the next stage can consume.

    \item \textbf{Traverse the call graph.}
    \textit{Entry points and chain construction.} An entry point is a call-graph vertex with no incoming edges --- a candidate place where execution can begin. From each entry point we walk the graph and emit a chain of function calls in the order they would execute at run time. The literal-assignment events recorded in stage 1 are inserted into the chain at the points where they occur, so the model sees the values a variable may carry by the time it reaches a call.
    \textit{Cycle handling and file-level scope.} Recursive and cyclic calls are excluded so that traversal terminates on graphs with arbitrary mutual-call structure. Direct self-calls (caller and callee coincide) are dropped during graph preparation, and returns to functions already present in the current chain are skipped. In the final configuration the traversal is also scoped to a single file: for a given entry point, only calls that stay within that file are kept. This matches our shift from package-level to file-level classification (see Section~\ref{sec:experiments}) --- each file gets its own textual description, and the package-level label is obtained by aggregating per-file predictions.

    \item \textbf{Render the chains as text.}
    Each call in a chain is mapped to an activity type (file-system access, network call, process creation, and so on). Calls that fall outside the activity classes of interest are not dropped --- they are collapsed into \emph{padding tokens}, single tokens of the form ``few / some / many padding activities'' that summarize a run of unclassified calls (ranges 6--19, 20--50, and 51+; runs of ${\leq}5$ unclassified calls are discarded). This lets the model see how sparse the relevant activities are in the chain without inflating the context with bookkeeping tokens.
    The chains are then ordered by when they may execute --- install-time first, then module-import-time, then the rest --- and concatenated into a single document.

    \item \textbf{Classify the text with BERT.}
    The document is fed to a BERT-based classifier, which returns a binary label: malicious or benign.

    \item \textbf{Explain malicious predictions.}
    When a sample is flagged as malicious, SHAP is applied to its textual description with a \emph{masker} (the SHAP component that decides which tokens to perturb), configured to split the text along activity boundaries. Each activity in the chain therefore becomes a single SHAP token whose contribution to the prediction can be scored.
    We compute each activity's contribution to the malicious class as the difference between the two classes' SHAP values:
    \begin{equation}
        s_i = \mathrm{shap}_i(\text{mal}) - \mathrm{shap}_i(\text{benign}).
    \end{equation}
    Service tokens (entry-point boundaries and padding tokens) are excluded from the ranking.
    We return the top-$k$ activities with positive contribution ($s_i > 0$); if fewer than $k$ are positive, fewer are returned. The rationale is discussed in Section~\ref{sec:explainability}. Each selected activity is reported with its location in code (file and line), letting the user jump straight to the snippet that drove the decision.
\end{enumerate}

The end result is a per-package verdict accompanied, for malicious cases, by a ranked list of suspicious activities pinpointed to file and line. The next section describes the dataset on which this pipeline is trained and evaluated.

\subsection{Dataset}
\label{sec:dataset}

The training data was assembled from several complementary sources covering both benign and malicious packages across the PyPI and npm ecosystems.

\textbf{Benign packages.} We collected clean packages from the following sources:
\begin{itemize}
    \item PyPI TOP-4000 --- the 4000 most-downloaded projects from PyPI.
    \item npm TOP-5000 --- the 5000 most-downloaded projects from npm.
\end{itemize}

To filter for benign data, we relied on the assumption that TOP-N packages, when subjected to a two-month quarantine period following their release date, can be treated as clean by default and are unlikely to contain malicious activity.

\textbf{Malicious packages.} Malicious samples were drawn from:
\begin{itemize}
    \item PyPI Malregistry and Datadog --- public datasets of projects in which malicious code has been confirmed.
    \item Malicious PyPI and npm packages curated by the PT Supply Chain Security expertise team.
\end{itemize}

\textbf{Preprocessing.} The raw data was filtered to remove projects containing no Python or JS code, as well as duplicate entries (since some malicious projects differed only by version).

The resulting package-level dataset composition is shown in Table~\ref{tab:dataset_packages}:

\begin{table}[h]
\centering
\caption{Package-level dataset composition.}
\label{tab:dataset_packages}
\begin{tabular}{lll}
\toprule
\textbf{Language} & \textbf{Train set} & \textbf{Validation set} \\
\midrule
Python     & 6088 pkgs (3047 mal.)   & 1779 pkgs (1018 mal.) \\
JavaScript & 10334 pkgs (6697 mal.)  & 2585 pkgs (1675 mal.) \\
\bottomrule
\end{tabular}
\end{table}

\subsection{File-Level Labeling}
\label{sec:filelevel}

The composition above is reported per package, but MOLOT actually classifies individual files. To work around the model's limited context window, we classify individual project files rather than whole packages. File-level models need file-level labels, so our experts hand-labeled 373 Python files and 21 JS files.

We turned these expert labels into a benchmark for measuring LLM labeling quality. The benchmark scores how well an LLM agrees with experts at two granularities --- lines of code and malicious-activity classes --- and tracks file-level accuracy and systematic errors. See Appendix~\ref{app:filelevel} for the full results.

The final activity-sequence dataset comes out of call-graph extraction and NLP preprocessing. The ``Train/Validation set'' columns count only the files that yielded at least one activity sequence --- the input the BERT classifier actually sees. The ``Validation: all files'' column gives the total file count in the validation set (matching the Support column in our file-level comparison against open-source tools); at inference time we label any file without an extracted sequence as ``benign''.

\begin{table}[h]
\centering
\caption{Activity-sequence dataset composition (file-level).}
\label{tab:dataset_sequences}
\setlength{\tabcolsep}{3pt}
\footnotesize
\begin{tabular}{lccc}
\toprule
\textbf{Language} & \makecell{\textbf{Train set}\\\textbf{(sequences)}} & \makecell{\textbf{Val set}\\\textbf{(sequences)}} & \makecell{\textbf{Val:}\\\textbf{all files}} \\
\midrule
Python     & 47221 (3940 mal.) & 11502 (1269 mal.) & 41579 \\
JavaScript & 18902 (6909 mal.) & 4740 (1862 mal.)  & 32292 \\
\bottomrule
\end{tabular}
\end{table}

\subsection{Public Benchmark}
\label{sec:publicbench}

Having defined the activity-sequence dataset, we carve a public subset out of its validation split. We release this subset as an open benchmark. The repository ships a class-balanced dataset, file- and package-level labels, and scripts for running formal-analysis tools; Table~\ref{tab:public_bench} summarizes its composition. The next section shows empirically that metrics on this public benchmark track those on our full private validation set.

\begin{table}[h]
\centering
\caption{Public benchmark composition.}
\label{tab:public_bench}
\begin{tabular}{lc}
\toprule
\textbf{Language} & \textbf{Public benchmark} \\
\midrule
Python     & 400 (200 malicious) \\
JavaScript & 400 (200 malicious) \\
\bottomrule
\end{tabular}
\end{table}

\section{Experiment Results}
\label{sec:experiments}

We first isolate the representation changes that drive MOLOT's quality, then study cross-language transfer, benchmark against open-source tools, and finally measure explainability, runtime, and real-world behavior.

\textbf{A note on the datasets.} We report metrics under two complementary scenarios that isolate the contribution of each component. The first evaluates the \emph{model-only evaluation}: metrics are computed on the validation split of the activity-sequence dataset (see Table~\ref{tab:dataset_sequences}), which contains only files from which an activity sequence was successfully extracted. This is the setting used for the configuration and language-transfer experiments (Sections~\ref{sec:key_mods} and~\ref{sec:language}; Tables~\ref{tab:pkg_features}, \ref{tab:file_features}, \ref{tab:lang_pkg}, and~\ref{tab:lang_file}), and it measures classifier quality independently of extraction coverage. The second is an \emph{end-to-end} evaluation of the complete pipeline, as it runs in a CI/moderation deployment: in Sections~\ref{sec:opensource} and~\ref{sec:publicresults} (Tables~\ref{tab:opensource_pkg}, \ref{tab:opensource_file}, \ref{tab:public_pkg}, and~\ref{tab:public_file}), metrics span every file in a package --- including files with no extracted sequence, which the model labels benign by default. A third axis, the coverage of the extraction frontend (how many files yield activity chains), would in principle deserve its own measurement; because the same frontend is used in every experiment, it stays fixed and cannot confound the comparison. The two scenarios therefore give a complete picture --- the classifier on its own, and the combined extraction-plus-classification pipeline. The gap between them also explains the lower Recall on the external benchmark relative to internal validation: malicious files with no extracted sequence never enter the positive predictions.

\subsection{Key Performance-Enhancing Modifications}
\label{sec:key_mods}

Our starting point was CEREBRO --- a behavioral classifier that operates on a textual description of call chains. Reproducing it surfaced two critical issues. First, the textual description for large packages did not fit into the model's context window. Second, the description embedded file and function names, which the model latched onto as shortcuts instead of learning the underlying code behavior.

We also observed that most malicious packages place their payload at the very beginning of a file, so in practice the original CEREBRO suffered relatively little from the context limit. That property, however, is an artifact of past campaigns and cannot be assumed to hold in future ones. The modifications below address both shortcomings; the supporting experiments are documented in Appendix~\ref{app:leakage}.

\begin{itemize}
    \item \textbf{Per-file classification.} To stay within the context budget, we switched from package-level to file-level classification; the package prediction is obtained by aggregating per-file labels using an OR policy --- a package is labeled malicious if at least one of its files is classified as malicious.
    \item \textbf{No file or entry-point names.} To curb overfitting, we removed file names and entry-point names from the activity chains. The model had been relying on keywords in entry-point names rather than on the activities themselves (see Appendix~\ref{app:leakage}).
    \item \textbf{Textual representation.} We revised how activity chains are rendered as text: call arguments that carry behavioral information are preserved to make the sequences more diverse, while calls outside the activity classes of interest are compressed into padding tokens rather than dropped. Together, these changes give the model richer behavioral content and a signal about the sparsity and distribution of security-relevant activities in the code.
\end{itemize}

The textual representation of code exposes two configurable features, which can be enabled independently:
\begin{itemize}
    \item \textbf{paddings} --- calls that fall outside the activity classes of interest are replaced with padding tokens rather than dropped (see step~3 in Section~\ref{sec:approach}). This lets the model use the positional structure of the chain (M1 configuration).
    \item \textbf{args} --- next to each activity in the text representation, we inline the literal values of arguments and of variables assigned to the call's result, whenever they may be passed into the call (M2 configuration).
\end{itemize}

Table~\ref{tab:pkg_features} evaluates both features independently and in combination on top of the M0 baseline. All variants are trained on a joint JS+Python dataset so that per-language metrics remain directly comparable.

\begin{table*}[tb]
\centering
\caption{Package-level metrics under different feature configurations. Model-only evaluation.}
\label{tab:pkg_features}
\footnotesize
\setlength{\tabcolsep}{4pt}
\begin{tabular}{lcccccccc}
\toprule
& & & \multicolumn{3}{c}{\textbf{JavaScript}} & \multicolumn{3}{c}{\textbf{Python}} \\
\cmidrule(lr){4-6} \cmidrule(lr){7-9}
\textbf{Configuration} & \textbf{args} & \textbf{paddings} & \textbf{Precision} & \textbf{Recall} & \textbf{F1} & \textbf{Precision} & \textbf{Recall} & \textbf{F1} \\
\midrule
CEREBRO --- original$^\dagger$          & ---       & ---       & 1.00  & 1.00  & 1.00            & 0.99  & 0.96  & 0.97 \\
M0 --- baseline                         & ---       & ---       & 0.930 & 0.966 & 0.948           & 0.845 & 0.928 & 0.885 \\
M1 --- + paddings                       & ---       & \checkmark & 0.951 & 0.962 & 0.957           & 0.889 & 0.925 & 0.907 \\
M2 --- + args                           & \checkmark & ---       & 0.946 & 0.964 & 0.955           & 0.884 & 0.942 & 0.912 \\
M3 --- + args \& paddings (final)       & \checkmark & \checkmark & 0.964 & 0.969 & \textbf{0.966}  & 0.922 & 0.929 & \textbf{0.926} \\
\midrule
\multicolumn{9}{l}{%
  \footnotesize$^\dagger$Includes file and function names in activity chains (identifier leakage); metrics are not directly comparable with M0--M3.%
} \\
\bottomrule
\end{tabular}
\end{table*}

Although M0--M3 operate at the file level, package predictions are obtained by aggregating file labels. In JS, F1 therefore improves steadily at both granularities: package-level F1 rises from 0.948 to 0.966, while file-level F1 increases from 0.865 to 0.888. The package-level gain is moderate because the baseline already identifies most malicious packages correctly; the file-level improvement shows that the refined representation captures a larger fraction of malicious chains within those packages.

For Python, the improvements are driven primarily by reduced false positives. Cleaner chain structure raises file-level precision from 0.719 to 0.875, which in turn increases package-level precision from 0.845 to 0.922.

\begin{table*}[tb]
\centering
\caption{File-level metrics under different feature configurations. Model-only evaluation.}
\label{tab:file_features}
\footnotesize
\setlength{\tabcolsep}{4pt}
\begin{tabular}{lcccccccc}
\toprule
& & & \multicolumn{3}{c}{\textbf{JavaScript}} & \multicolumn{3}{c}{\textbf{Python}} \\
\cmidrule(lr){4-6} \cmidrule(lr){7-9}
\textbf{Configuration} & \textbf{args} & \textbf{paddings} & \textbf{Precision} & \textbf{Recall} & \textbf{F1} & \textbf{Precision} & \textbf{Recall} & \textbf{F1} \\
\midrule
M0 --- baseline                  & ---       & ---       & 0.897 & 0.835 & 0.865           & 0.719 & 0.906 & 0.795 \\
M1 --- + paddings                & ---       & \checkmark & 0.925 & 0.840 & 0.880           & 0.810 & 0.906 & 0.856 \\
M2 --- + args                    & \checkmark & ---       & 0.915 & 0.849 & 0.881           & 0.818 & 0.904 & 0.859 \\
M3 --- + args \& paddings (final)& \checkmark & \checkmark & 0.922 & 0.857 & \textbf{0.888}  & 0.875 & 0.904 & \textbf{0.889} \\
\bottomrule
\end{tabular}
\end{table*}

As Table~\ref{tab:file_features} shows, padding tokens benefit Python substantially more than JS (file-level F1: 0.795 $\rightarrow$ 0.856 versus 0.865 $\rightarrow$ 0.880). Python call chains tend to be longer and noisier, so explicit gaps help the model recover their structural organization.

Arguments provide a comparable improvement on Python and a smaller but consistent recall gain on JS.

We adopt the combination of both features as the final configuration. At the file level, it achieves the strongest F1 score in both ecosystems (JS 0.888, Py 0.889), while also improving package-level performance (JS 0.966, Py 0.926). Across both ecosystems, M3 is the strongest configuration among the variants we evaluated.

\subsection{Language-Specific Results}
\label{sec:language}

Having fixed the M3 representation, we next ask whether a single model can serve both ecosystems or whether each language needs its own.

Tables~\ref{tab:lang_pkg} and~\ref{tab:lang_file} show how the composition of the training set (\texttt{python + javascript} / \texttt{javascript} / \texttt{python}) affects model quality on the validation splits of both languages. All experiments use the best feature configuration (\texttt{arguments + paddings}).

\begin{table*}[tb]
\centering
\caption{Package-level metrics by training-set composition. Model-only evaluation.}
\label{tab:lang_pkg}
\footnotesize
\setlength{\tabcolsep}{4pt}
\begin{tabular}{lcccccc}
\toprule
& \multicolumn{3}{c}{\textbf{JavaScript}} & \multicolumn{3}{c}{\textbf{Python}} \\
\cmidrule(lr){2-4} \cmidrule(lr){5-7}
\textbf{Trained on} & \textbf{Precision} & \textbf{Recall} & \textbf{F1} & \textbf{Precision} & \textbf{Recall} & \textbf{F1} \\
\midrule
python + javascript & 0.964 & 0.969 & \textbf{0.966} & 0.922 & 0.929 & 0.926          \\
javascript          & 0.956 & 0.978 & \textbf{0.967} & 0.761 & 0.865 & 0.810          \\
python              & 0.932 & 0.331 & 0.488          & 0.960 & 0.915 & \textbf{0.937} \\
\bottomrule
\end{tabular}

\vspace{0.9em}

\caption{File-level metrics by training-set composition. Model-only evaluation.}
\label{tab:lang_file}
\footnotesize
\setlength{\tabcolsep}{4pt}
\begin{tabular}{lcccccc}
\toprule
& \multicolumn{3}{c}{\textbf{JavaScript}} & \multicolumn{3}{c}{\textbf{Python}} \\
\cmidrule(lr){2-4} \cmidrule(lr){5-7}
\textbf{Trained on} & \textbf{Precision} & \textbf{Recall} & \textbf{F1} & \textbf{Precision} & \textbf{Recall} & \textbf{F1} \\
\midrule
python + javascript & 0.922 & 0.857 & \textbf{0.888} & 0.875 & 0.904 & 0.889          \\
javascript          & 0.903 & 0.868 & 0.886          & 0.462 & 0.742 & 0.570          \\
python              & 0.871 & 0.222 & 0.354          & 0.958 & 0.889 & \textbf{0.922} \\
\bottomrule
\end{tabular}
\end{table*}

Training on a single language gives strong performance on the native language but drops sharply on the other: the JS-only model reaches just 0.570 F1 (file) on Python, and the Python-only model reaches just 0.354 F1 (file) on JS.

The drop is not symmetric across training sets. The JS-trained model retains reasonable performance on Python (file-F1 0.570, pkg-F1 0.810), suggesting that the malicious-behavior chains in our JS training data largely subsume those seen in Python. The reverse does not hold: the Python-trained model collapses on JS, recovering only 0.331 of malicious JS packages (pkg-F1 0.488). This indicates that our JS training set covers additional malicious-behavior chains that are underrepresented in the Python data, rather than any inherent property of the languages themselves.

Joint training nearly matches the specialized models on their native language: on JS it nearly matches the JS-only model (pkg-F1 0.966 vs 0.967), and on Python it remains close to the Python-only model (pkg-F1 0.926 vs 0.937), while maintaining high quality on both languages. The small remaining gap is likely due to the imbalance and greater diversity of JS samples in the training set.

Most misclassifications came from projects with obfuscated code. In those cases, not every call of interest was extracted correctly, causing the model to lose important signal.

These results allow us to use a single jointly trained model in the final system for both JS and Python.

\subsection{Comparison with Open-Source Tools}
\label{sec:opensource}

With the jointly trained model fixed, we now position it against existing detectors. To evaluate the model against existing open-source malicious code detection tools, we conducted a benchmark on the validation dataset.

The following tools were compared:
\begin{itemize}
    \item \textbf{MOLOT} --- our solution combining SAST call-graph analysis with a BERT-based model.
    \item \textbf{Microsoft OSSGadget} (\texttt{oss-detect-backdoor}) --- executed via Microsoft AppInspector using the BackdoorRules rule set~\cite{ossgadget,appinspector}.
    \item \textbf{bandit4mal} --- a fork of Bandit with rules for malicious code detection (Python only)~\cite{bandit4mal,vu2022benchmark}.
    \item \textbf{semgrep} using the rules from \href{https://github.com/apiiro/malicious-code-ruleset}{apiiro/malicious-code-ruleset}~\cite{semgrep,apiiro_ruleset}.
\end{itemize}

Each applicable tool was executed on unpacked package archives from the validation dataset. Predictions were collected for both packages and individual files, and precision, recall, and F1 scores were computed at both granularity levels.

For all tools, files processed successfully but not flagged by any rule were treated as benign. This enables a consistent comparison at the file level. File-level support differs across tools because some scanners only process specific file types or may skip files that cannot be parsed successfully.

\begin{table*}[tb]
\centering
\caption{Package-level results on the full validation dataset (with 95\% bootstrap CIs). End-to-end evaluation.}
\label{tab:opensource_pkg}
\footnotesize
\setlength{\tabcolsep}{3pt}
\begin{tabular}{llcccc}
\toprule
\textbf{Tool} & \textbf{Eco.} & \textbf{Precision} & \textbf{Recall} & \textbf{F1} & \textbf{Support} \\
\midrule
MOLOT (ours) & Py & \valci{0.924}{0.905, 0.940} & \valci{0.923}{0.906, 0.938} & \valci{\textbf{0.923}}{0.911, 0.935} & 1579 \\
ossgadget    & Py & \valci{0.669}{0.642, 0.696} & \valci{0.839}{0.815, 0.861} & \valci{0.744}{0.723, 0.764} & 1579 \\
bandit4mal   & Py & \valci{0.554}{0.527, 0.583} & \valci{0.669}{0.639, 0.697} & \valci{0.606}{0.581, 0.630} & 1579 \\
semgrep      & Py & \valci{0.599}{0.568, 0.632} & \valci{0.518}{0.486, 0.550} & \valci{0.556}{0.529, 0.584} & 1579 \\
\midrule
MOLOT (ours) & JS & \valci{0.964}{0.949, 0.977} & \valci{0.908}{0.888, 0.927} & \valci{\textbf{0.935}}{0.921, 0.948} & 1004 \\
ossgadget    & JS & \valci{0.831}{0.801, 0.858} & \valci{0.747}{0.718, 0.776} & \valci{0.787}{0.763, 0.808} & 1004 \\
semgrep      & JS & \valci{0.861}{0.831, 0.890} & \valci{0.604}{0.565, 0.643} & \valci{0.710}{0.679, 0.740} & 1004 \\
\bottomrule
\end{tabular}

\vspace{0.9em}

\caption{File-level results on the full validation dataset (with 95\% bootstrap CIs). End-to-end evaluation.}
\label{tab:opensource_file}
\footnotesize
\setlength{\tabcolsep}{3pt}
\begin{tabular}{llcccc}
\toprule
\textbf{Tool} & \textbf{Eco.} & \textbf{Precision} & \textbf{Recall} & \textbf{F1} & \textbf{Support} \\
\midrule
MOLOT (ours) & Py & \valci{0.846}{0.811, 0.876} & \valci{0.791}{0.762, 0.823} & \valci{\textbf{0.818}}{0.794, 0.843} & 41579 \\
ossgadget    & Py & \valci{0.161}{0.115, 0.223} & \valci{0.664}{0.624, 0.706} & \valci{0.260}{0.197, 0.334} & 48689 \\
bandit4mal   & Py & \valci{0.126}{0.100, 0.155} & \valci{0.636}{0.609, 0.666} & \valci{0.210}{0.172, 0.250} & 47904 \\
semgrep      & Py & \valci{0.163}{0.104, 0.261} & \valci{0.444}{0.414, 0.474} & \valci{0.239}{0.169, 0.330} & 43664 \\
\midrule
MOLOT (ours) & JS & \valci{0.908}{0.880, 0.931} & \valci{0.774}{0.727, 0.823} & \valci{\textbf{0.835}}{0.802, 0.869} & 32292 \\
ossgadget    & JS & \valci{0.194}{0.127, 0.326} & \valci{0.479}{0.447, 0.513} & \valci{0.277}{0.199, 0.393} & 32292 \\
semgrep      & JS & \valci{0.564}{0.389, 0.720} & \valci{0.344}{0.309, 0.385} & \valci{0.428}{0.359, 0.491} & 31746 \\
\bottomrule
\end{tabular}
\end{table*}

Confidence intervals were estimated using non-parametric bootstrap resampling (1000 resamples, 95\% percentile intervals). At the package level, packages were resampled with replacement for each bootstrap iteration. At the file level, we applied cluster bootstrap resampling by package: each iteration sampled packages with replacement and included all files belonging to the selected packages. This procedure accounts for dependencies between files within the same package (e.g., shared author, coding style, or malicious campaign) and avoids the overly narrow confidence intervals produced by naive row-wise resampling.

Overall, MOLOT achieves the highest F1 scores among the evaluated tools at both the package and file levels in both ecosystems (Tables~\ref{tab:opensource_pkg} and~\ref{tab:opensource_file}).

\subsection{Public Benchmark Results}
\label{sec:publicresults}

The comparison above runs on our private validation set; we now repeat it on the public subset so others can reproduce these numbers. Since the full validation dataset cannot be released publicly, we provide a class-balanced subset of 800 packages (400 Python and 400 JS packages, with a 50/50 benign-to-malicious ratio). This balanced subset is intended solely for reproducibility and comparative benchmarking. MOLOT and the same set of open-source tools were evaluated on this subset; Tables~\ref{tab:public_pkg} and~\ref{tab:public_file} report the package- and file-level results. The resulting metrics remain within the confidence intervals observed on the full private validation set (Tables~\ref{tab:opensource_pkg} and~\ref{tab:opensource_file}).

Benchmark scripts and resulting artifacts (\texttt{predictions.parquet}, \texttt{metrics.csv}, and \texttt{metrics\_ci.csv}) are publicly available together with the dataset repository (see Section~\ref{sec:data}).

\begin{table*}[tb]
\centering
\caption{Package-level results on the public benchmark subset (with 95\% bootstrap CIs). End-to-end evaluation.}
\label{tab:public_pkg}
\footnotesize
\setlength{\tabcolsep}{3pt}
\begin{tabular}{llcccc}
\toprule
\textbf{Tool} & \textbf{Eco.} & \textbf{Precision} & \textbf{Recall} & \textbf{F1} & \textbf{Support} \\
\midrule
MOLOT (ours) & Py & \valci{0.896}{0.855, 0.936} & \valci{0.865}{0.816, 0.909} & \valci{\textbf{0.880}}{0.845, 0.911} & 400 \\
ossgadget    & Py & \valci{0.554}{0.496, 0.609} & \valci{0.770}{0.713, 0.824} & \valci{0.644}{0.597, 0.690} & 400 \\
bandit4mal   & Py & \valci{0.519}{0.467, 0.568} & \valci{0.885}{0.838, 0.925} & \valci{0.654}{0.607, 0.696} & 400 \\
semgrep      & Py & \valci{0.557}{0.496, 0.617} & \valci{0.685}{0.620, 0.748} & \valci{0.614}{0.560, 0.664} & 400 \\
\midrule
MOLOT (ours) & JS & \valci{0.912}{0.869, 0.952} & \valci{0.885}{0.840, 0.923} & \valci{\textbf{0.898}}{0.866, 0.927} & 400 \\
ossgadget    & JS & \valci{0.671}{0.607, 0.728} & \valci{0.805}{0.746, 0.854} & \valci{0.732}{0.682, 0.776} & 400 \\
semgrep      & JS & \valci{0.746}{0.678, 0.809} & \valci{0.645}{0.576, 0.714} & \valci{0.692}{0.637, 0.742} & 400 \\
\bottomrule
\end{tabular}

\vspace{1.4em}

\caption{File-level results on the public benchmark subset (with 95\% bootstrap CIs). End-to-end evaluation.}
\label{tab:public_file}
\footnotesize
\setlength{\tabcolsep}{3pt}
\begin{tabular}{llcccc}
\toprule
\textbf{Tool} & \textbf{Eco.} & \textbf{Precision} & \textbf{Recall} & \textbf{F1} & \textbf{Support} \\
\midrule
MOLOT (ours) & Py & \valci{0.849}{0.798, 0.903} & \valci{0.675}{0.625, 0.727} & \valci{\textbf{0.752}}{0.709, 0.796} & 10203 \\
ossgadget    & Py & \valci{0.183}{0.136, 0.248} & \valci{0.566}{0.506, 0.633} & \valci{0.277}{0.218, 0.345} & 10203 \\
bandit4mal   & Py & \valci{0.160}{0.119, 0.212} & \valci{0.742}{0.686, 0.801} & \valci{0.263}{0.205, 0.331} & 10203 \\
semgrep      & Py & \valci{0.252}{0.195, 0.327} & \valci{0.513}{0.457, 0.583} & \valci{0.338}{0.279, 0.403} & 10167 \\
\midrule
MOLOT (ours) & JS & \valci{0.838}{0.755, 0.907} & \valci{0.721}{0.633, 0.810} & \valci{\textbf{0.775}}{0.704, 0.843} & 15305 \\
ossgadget    & JS & \valci{0.197}{0.112, 0.345} & \valci{0.632}{0.563, 0.711} & \valci{0.300}{0.190, 0.451} & 15305 \\
semgrep      & JS & \valci{0.567}{0.475, 0.664} & \valci{0.418}{0.342, 0.504} & \valci{0.481}{0.413, 0.553} & 14913 \\
\bottomrule
\end{tabular}
\end{table*}

Confidence intervals were estimated using the same methodology as in Section~\ref{sec:opensource}. The metrics obtained for the open-source tools on the public subset remain within the confidence intervals observed on the full private validation set (Tables~\ref{tab:opensource_pkg} and~\ref{tab:opensource_file}).

File-level support differs across tools because each scanner processes a different set of file extensions (e.g., OSSGadget scans \texttt{.ts} files, whereas semgrep does not). In addition, semgrep reached the 600-second timeout limit on five packages and therefore did not produce file-level outputs for them.

\subsection{Explainability}
\label{sec:explainability}

Beyond raw detection quality, MOLOT must justify its verdicts, so we next evaluate how well its SHAP-based explanations localize malicious code. The objective of the experiment was to identify the activity selection strategy that most closely matches expert annotations of suspicious code lines. To this end, we used a subset of 344 malicious package files. For each file, an expert annotated the set of lines containing the core malicious behavior. Algorithm performance was evaluated using intersection over union (IoU), precision, and recall between the set of lines associated with the selected activities and the expert annotations.

A preliminary observation is that the average overlap between the complete activity chain and the expert annotations is 0.75 (median: 0.93). This indicates that call graph traversal alone already provides strong localization of the malicious code fragment. Consequently, the role of SHAP-based selection is primarily to reduce the number of reported activities to the most influential ones while preserving high recall.

The following activity selection strategies were evaluated:
\begin{itemize}
    \item Top-$k$ activities with the highest contribution;
    \item Top-$k$ activities restricted to those with positive contribution;
    \item All activities with positive contribution;
    \item Activities ranked by contribution whose cumulative positive contribution exceeds a threshold fraction of the total positive contribution mass.
\end{itemize}

Fixed-size top-$k$ strategies without sign filtering achieved the highest recall but showed reduced precision. In the absence of clearly suspicious activities, these methods still return $k$ elements, including neutral or weakly negative ones. Restricting selection to positively contributing activities consistently improved precision across all $k$ values with only a minor reduction in recall. In addition, this approach naturally adapts the output size: simple cases typically yield 1--2 activities, whereas more complex samples may produce up to $k$ activities.

Threshold-based strategies achieved the highest precision; however, they exhibited lower recall and frequently returned only a single activity, which limits their practical usefulness for analysts. Selecting all positively contributing activities maximized recall, but resulted in highly variable output sizes that occasionally reached dozens of activities.

Based on these observations, the final configuration employs \textbf{top-$k$ selection over positively contributing activities}. Compared to the standard top-$k$ strategy, this approach improves precision, maintains high recall, and constrains the output size, thereby improving analyst usability.

\subsection{Performance Evaluation}
\label{sec:performance}

\begin{table*}[tb]
\centering
\caption{Per-package processing time and memory consumption across pipeline stages.}
\label{tab:performance}
\footnotesize
\setlength{\tabcolsep}{4pt}
\begin{tabular}{llrrrll}
\toprule
\textbf{Step} & \textbf{Device} & \textbf{Avg. time} & \textbf{Median time} & \textbf{Max time} & \textbf{Slowdown vs.\ base} & \textbf{RAM} \\
\midrule
\multirow{3}{*}{SAST --- call-graph extraction}
    & CPU (2 cores) & 16.08\,s  & 14.67\,s  & 182.33\,s & $\times$1.24       & \multirow{3}{*}{1.10\,GB} \\
    & CPU (4 cores) & 13.31\,s  & 13.09\,s  & 127.57\,s & $\times$1.03       & \\
    & CPU           & 12.98\,s  & 13.39\,s  & 89.68\,s  & $\times$1 --- base & \\
\midrule
Text description generation & CPU & 6.76\,s & 3.94\,s & 18.95\,s & $\times$1 --- base & 400\,MB \\
\midrule
\multirow{4}{*}{BERT inference}
    & CPU (2 cores) & 576\,ms & 583\,ms & 1083\,ms & $\times$19.86       & \multirow{3}{*}{500\,MB} \\
    & CPU (4 cores) & 242\,ms & 223\,ms & 426\,ms  & $\times$8.34        & \\
    & CPU           & 117\,ms & 116\,ms & 339\,ms  & $\times$4.03        & \\
    & CUDA          & 29\,ms  & 29\,ms  & 251\,ms  & $\times$1 --- base  & 620\,MB VRAM \\
\midrule
\multirow{4}{*}{Suspicious-call extraction (SHAP)}
    & CPU (2 cores) & 99.1\,s & 19.5\,s & 322.6\,s & $\times$19.06       & \multirow{3}{*}{1.1\,GB} \\
    & CPU (4 cores) & 47.3\,s & 9.7\,s  & 129.5\,s & $\times$9.10        & \\
    & CPU           & 20.6\,s & 3.6\,s  & 61.3\,s  & $\times$3.96        & \\
    & CUDA          & 5.2\,s  & 1.1\,s  & 15.9\,s  & $\times$1 --- base  & 1.5\,GB VRAM \\
\bottomrule
\end{tabular}
\end{table*}

Accurate, explainable verdicts are only useful if the pipeline is fast and light enough to deploy, which we measure next. Before adding our solution to the product, we needed to evaluate the pipeline's performance --- specifically, the processing time per Python/JS package and its RAM usage.

The evaluation consisted of the following steps:
\begin{enumerate}
    \item Extracting the call graph from the project using the SAST engine of PT Application Inspector.
    \item Generating a textual description of the package by traversing the call graph and mapping calls to behavioral activities.
    \item Running model inference (BERT).
    \item Extracting suspicious calls --- we use SHAP to explain the model's prediction and identify the calls that contributed most to it.
\end{enumerate}

The entire pipeline runs inside a Docker container, and we use cAdvisor to measure memory consumption at every stage.

We tested on 400 JS/Python projects. The sample is evenly split between JS and Python and spans a range of project sizes, from packages with only a handful of files to large ones with hundreds. This wide size distribution is heavily skewed toward small packages, which is why the median is noticeably lower than the mean for most steps --- a few very large projects pull the average up.

For the model, we used \texttt{bert-base-uncased} and ran inference through the \texttt{pipeline} API from the \texttt{transformers} library.

Memory consumption does not depend on the number of CPU cores, so RAM is reported once per step. Rows labeled \texttt{CPU} without an explicit core count use all 16 cores of the test machine. Table~\ref{tab:performance} reports per-package processing time and memory consumption across the pipeline stages.

A few things stand out in these numbers. On GPU, the full pipeline is dominated by call-graph extraction, which is CPU-bound and cannot be accelerated further. On CPU-only configurations, SHAP-based suspicious-call extraction becomes the clear bottleneck, scaling almost linearly with the inverse of available cores. BERT inference itself is fast enough on CPU that GPU acceleration is a nice-to-have rather than a requirement for the model step alone --- but it becomes essential once SHAP is added.

The measurements were taken on the following hardware:
\begin{itemize}
    \item \textbf{OS} --- Ubuntu 22.04.3 LTS x86\_64.
    \item \textbf{CPU} --- Intel Xeon (Icelake), 16 cores @ 1.995\,GHz.
    \item \textbf{GPU} --- NVIDIA Tesla T4, 16\,GB.
\end{itemize}

These results meet our product requirements.

\subsection{Real-World Usefulness Evaluation}
\label{sec:realworld}

Controlled measurements aside, the decisive test is behavior on live traffic. As a pilot deployment, we integrated the model into the company's pypi/npm moderation pipeline. The solution was deployed on the internal npm and PyPI mirror: each new package submitted by a developer was automatically scanned by our tool as part of moderation. Evaluation was therefore conducted not on a pre-collected sample, but on the live stream of packages submitted by company developers.

Over the 2-month pilot, the tool scanned 2{,}672 packages, with an average processing time of 80 seconds per package --- comparable to other tools commonly used in CI pipelines. Since the stream consists predominantly of benign packages, it also provides an estimate of the false positive rate under real-world conditions: 9.24\% at the package level and 1.9\% at the file level. All flagged packages were subsequently reviewed by security experts and confirmed as benign.

These results indicate that the solution is applicable to real moderation workflows: the package-level false positive rate does not impose a significant burden on reviewers, and per-package latency is consistent with existing CI tooling.

\section{Limitations and Future Work}
\label{sec:limitations}

Despite these results, MOLOT has several limitations that also define our roadmap.

A key direction for future work is extending MOLOT to additional programming languages and package ecosystems. The CEREBRO paper notes that its approach can be generalized to new ecosystems by implementing a feature extractor and a behavior-sequence generator for each target language. In MOLOT, the sequence generator is largely universal, but feature extraction still requires language-specific security expertise. As a result, support for a new language can be added without retraining the classifier; to reach the quality of mature formal-analysis tools, it is sufficient to add a dedicated set of activity classes for the new ecosystem.

Although MOLOT is well suited for SAST workflows, its current performance is not yet sufficient for high-throughput supply-chain security scenarios where hundreds of thousands of package releases may need to be analyzed every day. The main bottleneck is graph construction. Optimizing call-graph extraction and reducing the cost of package processing is therefore one of the main engineering directions for future work.

Another direction is improving entry-point analysis and exploring graph neural networks. The current LLM-assisted labeling pipeline does not yet provide line-level accuracy high enough to train a more granular model reliably. Developing an agent-based approach to labeling malicious projects could make it possible to obtain finer supervision and move toward models that reason not only over file behavior sequences, but also over the structure of the call graph itself.

Finally, attacker techniques continue to evolve, and MOLOT is expected to expand to more languages over time. For this reason, we plan to keep updating the public benchmark by adding more diverse malicious campaigns and new programming languages.

\section{Data Availability}
\label{sec:data}

The public benchmark dataset is available in the repository: \href{https://github.com/False-Positive-Community/open-malicious-code-benchmark}{False-Positive-Community/open-malicious-code-benchmark}.

\section{Conclusion}
\label{sec:conclusion}

MOLOT demonstrates that static behavior-sequence modeling can serve as a practical foundation for malicious-code detection in modern SAST workflows. Compared with formal-analysis tools, the system achieves strong detection quality while reducing overfitting, and its SHAP-based explanation stage links malicious predictions back to concrete suspicious activities and source-code locations. The resulting engine is suitable for on-premise deployment, supports CPU-compatible operation, and has already been integrated into PT Application Security SAST starting with release 6.0.0.

We expect this work to contribute to the broader application security community by making malicious-code detection more reproducible, explainable, and comparable across tools. The release of the open benchmark is an important step in that direction: it gives researchers and practitioners a shared basis for evaluating systems in this class, comparing them with MOLOT, and tracking their evolution over time.

\section*{Acknowledgment}

This work was developed and supported within the processes and practices of the Data Science \& ML department. We also want to acknowledge the contributions of the related teams involved in the project:
\begin{itemize}
    \item The PT AppSec Research team, for initiating the project and sharing their expertise in static code analysis.
    \item Alexander Chikaylo, Ekaterina Gorynya, Oleg Khabarov, and Valeriy Pushkar, for their invaluable expert annotation of the dataset.
    \item The PT Supply Chain Security team, for sharing their expertise in supply-chain attacks, providing the annotation platform, and contributing data to the project.
    \item The PT DevSecOps infrastructure team, for early testing of the functionality in the company's secure development workflows.
    \item The PT Application Inspector team, for comprehensive support in preparing the project for product release.
\end{itemize}

\bibliographystyle{IEEEtran}
\bibliography{references}

\appendices

\section{Leakage of File Identifiers in Early Activity Chains}
\label{app:leakage}

\textbf{Symptom.} In early versions of the pipeline, activity chains contained the entrypoint identifier --- specifically, the file name and the function name from which call-graph traversal began. The model learned the distribution of these identifiers from the training set and relied on particular substrings in file paths: substrings that frequently appeared in the benign portion of the dataset (\texttt{session}, \texttt{init}, \texttt{setup}, \ldots) biased predictions toward \texttt{benign}, producing false negatives on malicious packages with such names; substrings common in the malicious portion (\texttt{cse}, \texttt{encrypt}, \texttt{s3}, \ldots) had the opposite effect, causing false positives on benign packages. As a result, behavioral activities themselves could become secondary signals.

\begin{figure}[h]
    \centering
    \includegraphics[width=\columnwidth]{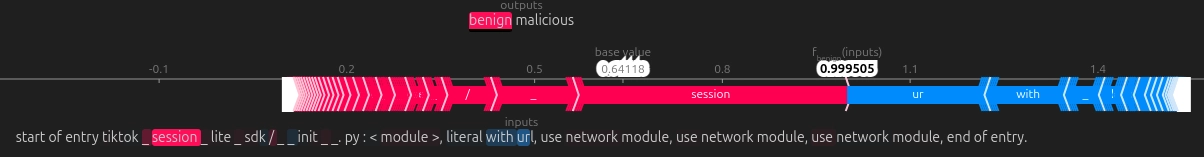}
    \caption{SHAP attribution before retraining: the substring \texttt{session} from the file name dominates the contribution toward the benign class.}
    \label{fig:shap_before}
\end{figure}

\textbf{Illustrative example (before/after).} A malicious package named \texttt{tiktok\_session\_lite\_sdk} (a TikTok typosquat). The same entrypoint is shown under two preprocessing variants:

\smallskip
\noindent\textcolor{beforered}{\textbf{\textsf{Before}}}~\textemdash~original preprocessing, entrypoint name included:
\begin{lstlisting}[style=plainBefore]
start of entry (*tiktok _ session _ lite _ sdk/_ _ init _ _ . py :<module>*),
literal with url, use network module, use network module, use network module,
end of entry.
\end{lstlisting}

\noindent\textcolor{aftergreen}{\textbf{\textsf{After}}}~\textemdash~current preprocessing, entrypoint name removed:
\begin{lstlisting}[style=plainAfter]
start of entry, literal with url,
use network module, use network module, use network module,
end of entry.
\end{lstlisting}

\textbf{What SHAP reveals.} \emph{Before:} the model incorrectly classifies the package as \texttt{benign} with a confidence score of \textbf{0.9995} against a baseline of \textbf{0.64} --- a False Negative. The largest positive contribution toward the \texttt{benign} prediction comes from the token \texttt{session} extracted from the file name (see Fig.~\ref{fig:shap_before}). In the training distribution, the substring \texttt{session} is statistically associated with benign packages (TikTok session SDKs, web-session utilities, etc.), and its presence in the typosquat package name outweighs the network-related runtime activities.

\begin{figure}[h]
    \centering
    \includegraphics[width=\columnwidth]{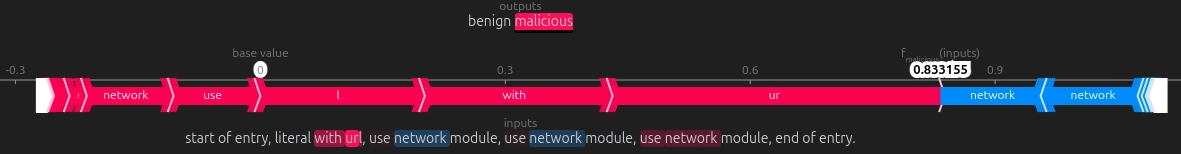}
    \caption{SHAP attribution after retraining: behavioral tokens (\texttt{network}, \texttt{url}, \texttt{use network module}) drive the malicious-class prediction.}
    \label{fig:shap_after}
\end{figure}

\emph{After:} the retrained model correctly predicts \texttt{malicious} with a score of \textbf{0.833} against a baseline of \textbf{0}. The strongest positive contributions now come from network-related activity tokens such as \texttt{network}, \texttt{url}, and \texttt{use network module} --- i.e., the activity sequence itself rather than lexical artifacts from the file name (see Fig.~\ref{fig:shap_after}). After retraining on activity chains without entrypoint identifiers, the model relies less on lexical file-name correlations and more on behavioral patterns. The SHAP baselines are therefore not directly comparable between the two variants.

\textbf{Notes.} The snippets are shown exactly as SHAP sees them after BERT tokenization (hence the spaces around separators such as \texttt{\_}, \texttt{/}, and \texttt{.} in the entrypoint name). Activity names correspond to natural-language descriptions of runtime behavior classes (\texttt{use network module}, \texttt{access file system}, \texttt{use data encoding module}, \texttt{use dynamic code generation}, \ldots); where applicable, argument labels are appended (e.g., \texttt{literal with url}, \texttt{\ldots with base64}).

\textbf{Conclusion.} Removing file and function names from activity chains --- followed by retraining on the updated representation --- is one of the key modifications described in Section~\ref{sec:key_mods}. This change reduces the model's dependence on lexical file-identifier correlations and encourages reliance on semantically meaningful behavioral features. Aggregate before/after metrics are reported in Section~\ref{sec:experiments} and are not repeated here.

\section{File-Level Labeling}
\label{app:filelevel}

\subsection*{LLM-assisted file-level annotation}

The transition from package-level to file-level classification requires labels at the same granularity as the model input. To construct such labels, we first prepared an expert-annotated seed set comprising 394 malicious files drawn from previously reviewed malicious packages. These annotations served two purposes: they provided a reference set for evaluating automated labeling quality, and they established the target taxonomy and localization standard used in the subsequent large-scale labeling stage.

We used the following metrics to quantify agreement between expert and LLM annotations:
\begin{itemize}
    \item \textbf{IOU class/lines, \%} --- intersection-over-union between expert and LLM annotations, requiring agreement both on the malicious-activity class (for example, STEALER, BACKDOOR, or DOWNLOADER) and on the corresponding source-code lines.
    \item \textbf{IOU lines, \%} --- intersection-over-union over source-code lines only, independent of activity class.
    \item \textbf{Lines with activities, \%} --- the percentage of snippets containing activities in which the LLM identifies potentially interesting calls.
    \item \textbf{Recall (file), \%} --- the proportion of expert-malicious files for which the LLM produced at least one snippet with an activity class other than NONE. This metric captures whether the model detects the presence of malicious behavior at the file level, independently of exact localization and class assignment.
    \item \textbf{Parsing and infrastructure errors} --- the number of responses that failed to satisfy the required JSON schema, together with infrastructure failures such as context-window overflow, rate limits, or service unavailability.
\end{itemize}

The model-selection procedure was conducted in two stages. In the first stage, we evaluated prompting strategies on a preliminary set of models and selected the most reliable annotation prompt. In the second stage, we fixed this prompt and compared a broader set of candidate models. This separation was necessary because prompt design had a stronger effect on annotation quality than model choice under a fixed prompt, as shown by the results below.

\textbf{Stage 1. Prompting strategy selection.} We evaluated three prompting strategies with Structured Output on Llama 3.3 and Mistral 2411: zero-shot prompting, few-shot prompting, and few-shot prompting augmented with activity-class descriptions (few-shot + desc). Table~\ref{tab:llm_prompts} reports the results.

\begin{table*}[tb]
\centering
\caption{Stage 1 --- LLM prompting strategies. Mean rows compare against the zero-shot baseline for each metric.}
\label{tab:llm_prompts}
\footnotesize
\setlength{\tabcolsep}{4pt}
\begin{tabular}{llcccc}
\toprule
\textbf{Prompt type} & \textbf{Model} & \textbf{IOU class/lines, \%} & \textbf{IOU lines, \%} & \textbf{Lines with activities, \%} & \textbf{Response parsing errors} \\
\midrule
zero-shot       & Llama 3.3      & 48.6 & 50.4 & 92.6 & 43 \\
zero-shot       & Mistral 2411   & 44.1 & 51.9 & 74.9 & 15 \\
\rowcolor{gray!15}
\textbf{zero-shot} & \textbf{Mean vs zero-shot} & \textbf{46.4 (base)} & \textbf{51.2 (base)} & \textbf{83.8 (base)} & \textbf{29.0 (base)} \\
few-shot        & Llama 3.3      & 49.1 & 50.2 & 92.1 & 1 \\
few-shot        & Mistral 2411   & 54.8 & 53.8 & 90.1 & 8 \\
\rowcolor{gray!15}
\textbf{few-shot} & \textbf{Mean vs zero-shot} & \textbf{52.0 (+12.1\%)} & \textbf{52.0 (+1.7\%)} & \textbf{91.1 (+8.8\%)} & \textbf{4.5 ($-$84.5\%)} \\
few-shot + desc & Llama 3.3      & 53.8 & 54.9 & 93.7 & 0 \\
few-shot + desc & Mistral 2411   & 57.8 & 58.0 & 94.0 & 4 \\
\rowcolor{gray!15}
\textbf{few-shot + desc} & \textbf{Mean vs zero-shot} & \textbf{55.8 (+20.4\%)} & \textbf{56.5 (+10.4\%)} & \textbf{93.9 (+12.1\%)} & \textbf{2.0 ($-$93.1\%)} \\
\bottomrule
\end{tabular}
\end{table*}

Across the evaluated models, adding examples and explicit activity-class descriptions improved annotation overlap, increased the share of snippets with detected activities, and substantially reduced schema-parsing failures. Relative changes in the mean rows are computed against the zero-shot mean for the corresponding metric; negative values for parsing errors indicate fewer failures than the zero-shot baseline. In particular, the mean Lines with activities score increased from 83.8\% under zero-shot prompting to 93.9\% under few-shot + desc, indicating more consistent identification of activity-bearing code regions. We therefore adopted \textbf{few-shot prompting with activity-class descriptions and Structured Output} as the fixed annotation prompt for the second stage.

\textbf{Stage 2. Model comparison under the fixed prompt.} Using the selected prompt, we evaluated a broader set of models, including GPT-OSS 120B, Devstral-Small, Claude Sonnet 4, and several lightweight models with at most 8B parameters. The evaluation used the same expert-annotated benchmark of approximately 400 files. Claude Sonnet 4 was included as a closed frontier-model reference point available at the time of evaluation in August 2025, providing an upper-bound estimate of LLM-assisted labeling quality under the fixed prompt. Because it does not satisfy the self-hosting requirement, it was used only as a reference and was not considered for production labeling pipeline. Table~\ref{tab:llm_models} reports the comparison.

\begin{table*}[tb]
\centering
\caption{Stage 2 --- Model comparison under the fixed few-shot + desc prompt.}
\label{tab:llm_models}
\footnotesize
\setlength{\tabcolsep}{3pt}
\begin{tabular}{lccccc}
\toprule
\textbf{Model} & \textbf{IOU class/lines, \%} & \textbf{IOU lines, \%} & \textbf{Recall (file), \%} & \textbf{Lines w/ activities, \%} & \textbf{Infra. / SO parsing errors} \\
\midrule
Claude Sonnet 4 \emph{(closed, upper estimate)} & 62.1 & 60.4 & 96.7 & 95.0 & 9 / 1   \\
GPT-OSS 120B           & 56.4 & 55.5 & 99.2 & 97.0 & 19 / 4  \\
Llama 3.3 70B          & 53.5 & 54.6 & 97.6 & 93.7 & 24 / 0  \\
Mistral-Large 2411     & 57.9 & 58.0 & 99.2 & 94.0 & 44 / 11 \\
Qwen 2.5 72B           & 50.8 & 53.1 & 97.8 & 92.0 & 34 / 4  \\
Devstral-Small 2505    & 56.9 & 57.7 & 97.0 & N/A  & 38 / 8  \\
Llama 3.1 8B           & 36.3 & 43.3 & 99.2 & 90.0 & 0 / 147 \\
Qwen2.5-Coder 7B       & 47.6 & 46.4 & 98.1 & 92.0 & 0 / 85  \\
Mistral 7B             & 41.8 & 43.4 & 97.8 & 92.0 & 0 / 119 \\
\bottomrule
\end{tabular}
\end{table*}

The dominant failure modes were context-window overflow and output-format instability. Context overflow was most visible for models with shorter context windows: Mistral-Large 2411, with a 65K-token window, produced approximately twice as many dropped samples as models with 131K-token windows, such as Llama 3.3 70B and GPT-OSS 120B. Output-format instability was concentrated among smaller models with at most 8B parameters when guided JSON was unavailable; common failures included markdown-wrapped responses, \texttt{<think>} blocks, and unstructured prose.

Based on these results, \textbf{Llama 3.3 70B} and \textbf{GPT-OSS 120B} were selected for production annotation. Both models are suitable for self-hosted deployment, support a 131K-token context window, and produced only 0--4 parsing failures on the expert benchmark. They also maintained high activity-coverage scores: 93.7\% for Llama 3.3 70B and 97.0\% for GPT-OSS 120B on Lines with activities, which indicates that both models consistently marked potentially relevant code regions rather than returning sparse annotations. We additionally measured false-positive rate on an independent set of \textbf{6{,}607 unique files extracted from trusted projects}, disjoint from the expert-labeled benchmark. For both selected models, FPR was below \textbf{1.5\%}, which we consider acceptable for generating supervision for the downstream BERT classifier, since the expected share of mislabeled benign files in the training set remains small.

The two selected models provide complementary operational advantages. Llama 3.3 70B achieved stronger localization quality, whereas GPT-OSS 120B offered higher throughput due to its MoE architecture, with approximately 5B active parameters. Claude Sonnet 4 achieved the highest line-level IOU, exceeding the selected models by approximately 2--5 percentage points, but was excluded from the production pipeline because it is a closed model and therefore does not satisfy the self-hosting requirement.

After finalizing the prompt and model configuration, we applied the annotation pipeline to the remaining unlabeled data: \textbf{6{,}607 unique files extracted from malicious PyPI packages} and \textbf{25{,}591 unique files extracted from malicious npm packages}. The resulting file-level labels were then used as supervision for training the activity-sequence classifier.

\end{document}